\documentstyle[12pt]{article}
\makeatletter                                                           %LMS

\def\section{\@startsection {section}{1}{\z@}{-1.5ex plus -.5ex
minus -.2ex}{1ex plus .2ex}{\large\bf}}                                 %LMS

\def\@thmcountersep{}                                                   %LMS

\long\def\@makecaption#1#2{\vskip 10pt
\setbox\@tempboxa\hbox{#1. #2}   %LMS
   \ifdim \wd\@tempboxa >\hsize   % IF longer than one line:
       #1. #2\par                 %   THEN set as ordinary paragraph.
     \else                        %   ELSE  center.                     %LMS
       \hbox to\hsize{\hfil\box\@tempboxa\hfil}                         %LMS
   \fi}                                                                 %LMS

\def\ps@headings{                                                       %LMS
 \def\@oddhead{\footnotesize\rm\hfill\runninghead\hfill}
 \def\@evenhead{\@oddhead}                                              %LMS
 \def\@oddfoot{\rm\hfill\thepage\hfill}\def\@evenfoot{\@oddfoot} }
\makeatother                                                            %LMS

\begin{document}

\title{Gauge (In)variance, Mass and Parity in \\ D=3 Revisited}{}{}

\def\runninghead{DESER :\quad GAUGE (IN)VARIANCE, MASS
AND PARITY IN D=3}
\author{
{\em S. Deser} \thanks{The Martin Fisher School of Physics, Brandeis
University, Waltham, MA 02254, USA.  This work was supported by the
National Science Foundation under grant \#PHY88--04561.}}

\date{} % no date wanted.                                              %LMS

\pagestyle{headings}                                                   %LMS
\flushbottom                                                           %LMS

\maketitle
\vspace{-10pt} % include according to taste.

\begin{abstract}
We analyze the degree of equivalence between abelian topologically
massive, gauge-invariant, vector or tensor parity doublets and their
explicitly massive, non-gauge, counterparts.
We establish equivalence of field equations by exploiting a generalized
Stueckelberg invariance of the gauge systems.  Although
the respective excitation spectra and induced source-source interactions
are essentially identical, there are also differences, most dramatic being
those between the Einstein limits of the interactions in the tensor case:
the doublets avoid the discontinuity (well-known from D=4) exhibited
by Pauli--Fierz theory.
\end{abstract}

% The next definitions are just used by the sample text. Don't include
%them.

%\renewcommand {\d} {\delta }
%\newcommand {\Ctt} {\hbox{$\CSA_{\,3\to2}$}}
%\newcommand {\Ckl} {\hbox{$\CSA_{k\to\ell}$}}
%\newcommand {\B}   {${\sl B}_2$}
%\newcommand {\U}   {${\sl U}_2$}
%\newcommand {\BB}  {{{\sl B}_2}}
%\newcommand {\UU}  {{{\sl U}_2}}
%\newcommand {\CSA} {{\sl CSA}}

\def\th{\theta} \def\TH{\Theta}
\def\pa{\partial}
\def\k{\kappa}
\def\g{\gamma} \def\G{\Gamma}
\def\a{\alpha}
\def\b{\beta}
\def\d{\delta} \def\D{\Delta}
\def\e{\epsilon}
\def\z{\zeta}
\def\k{\kappa}
\def\l{\lambda} \def\L{\Lambda}
\def\m{\mu}
\def\n{\nu}
\def\x{\chi}
\def\p{\pi} \def\P{\Pi}
\def\r{\rho}
\def\s{\sigma} \def\S{\Sigma}
\def\t{\tau}
\def\y{\upsilon} \def\Y{\Upsilon}
\def\o{\omega} \def\O{\Omega}

\vspace{.1in}

It is a pleasure to dedicate this work to Dieter Brill on the occasion of
his 60$^{\rm th}$ birthday.  I have learned much from him over the
years, not least during our old collaborations on general relativity.  I
hope he will be entertained by these considerations of related theories
in another dimension.

\section{Introduction} % 1

Perhaps the most paradoxical feature of topologically massive
(TM) theories \cite{001,002} is that their gauge invariance coexists with
the finite mass and single helicity, parity violating, character of their
excitations.  This phenomenon, common to vector (TME) and tensor
(TMG) models, is special to 2+1 dimensions because in higher (odd)
dimensions the operative Chern--Simons (CS) terms are of at least cubic
order in these fields and so do not affect their kinematics; only higher
rank tensors could acquire a topological mass there.
Equally surprisingly, TM doublets, with mass parameters of opposite
sign, are not only invariant under combined parity and field
interchanges\footnote{Doublets of pure (non-propagating) vector CS
models have also been studied \cite{003}.} but they are equivalent in
essential respects to non-gauge vector or tensor models with ordinary
mass terms: Both their excitation contents and the interactions they
induce among their sources are essentially identical, as originally
indicated in \cite{002}.

To be sure, other
mass generation mechanisms exist in gauge theories: in D=2, the
Schwinger model --- spinor electrodynamics --- yields a massive photon,
but not at tree level.  In D=4, in addition to the Higgs mechanisms,
there is one involving a mixed CS-like term bilinear in antisymmetric
tensor and vector fields \cite{004}.  However, none
function, as this one does, purely with a single gauge field.
It is also well-known that apparently different free-field theories can
have the same excitation spectrum.  Indeed, there even exist self-dual
formulations of TM theories \cite{005,006,007,008} involving a single
field, in which gauge invariance is hidden; however, their minimal
couplings to a given source are then inequivalent.

My purpose here is to review and make more explicit how there can be
coexistence between those equivalences and the differences in gauge
character, as well as to point out that traces of these differences
nevertheless remain.  On the equivalence side, we will see that the TM
field equations can be written in the non-gauge form because their
gauge
freedom can be hidden through a generalized Stueckelberg field
redefinition.  We
will also discuss three ways in which differences manifest themselves.
First, because gauge invariance implies the existence of Gauss
constraints, long-range potentials carrying information about the
sources survive in the massive doublets, but have no counterpart in the
non-gauge models.  Second, we will compare the massive excitation
spectra with those of the limiting massless theories and explain why
normally massive theories always differ from the latter, while TME (but
not TMG) is one example in which the number  of
excitations (if not their helicities) agree with massless theory.  Third,
the Einstein limits of the
tensor models differ dramatically: As is well-known \cite{009} in D=4,
massive tensors have components which do not decouple from the
sources in this limit and lead to source-source interactions different
from those of Einstein theory; TMG, on the other hand, will be shown
to avoid this discontinuity.
The present analysis is restricted to abelian models; their nonlinear
generalizations are not obviously amenable to these
considerations.\footnote{Their nonlinear parts fail to superpose, and the
respective sources cannot simultaneously be covariantly conserved with
respect to each gauge field and to their sum.}  Extension to their
supersymmetric generalizations \cite{001,010}, on the other hand, is
straightforward.

\section{Vector Theories}

Consider first a topologically massive vector doublet, which will be
compared with massive vector (Proca) theory.  The respective free
Lagrangians are, in {\small (-- + +)} signature,
\begin{eqnarray}%1
L_{\rm VD} = L_{\rm TME} \: (A^1_\mu , m) \!\! & + & \!\! L_{\rm
TME}
(A^2_\mu , -m) \: - \textstyle{\frac{1}{\sqrt{2}}} \: j^\mu
(A^1_\mu + A^2_\mu ) \; , \nonumber \\ [.1in]
L_{\rm TME} \: (A_\mu , m) & \equiv & - \textstyle{\frac{1}{4}} \:
F^2_{\mu\nu} (A) ~ + ~ \textstyle{\frac{1}{4}}\: m \:
\epsilon^{\alpha\mu\nu} A_\alpha \, F_{\mu\nu}(A)
\end{eqnarray}
and
\begin{equation}%2
L_P (A_\m ) = - \textstyle{\frac{1}{4}}\: F^2_{\mu\nu} -
\textstyle{\frac{1}{2}} \: m^2 A^2_\mu - j^\mu \, A_\mu \; .
\end{equation}

By gauge invariance, the current must be conserved in TME and the
comparison can only be made if, as we assume henceforth,
$\pa_\m j^\m = 0$.  Let us review what is known \cite{002}:
Invariance of $L_D$ under combined parity and field interchange
conjugations is manifest, since the parity transform is equivalent to a
change of sign of $m$, which is just cancelled by the interchange
$A^1_\m \leftrightarrow A^2_\m$.  Next, the equivalence of the
effective current-current interactions generated by (1) and by (2) follows
from the forms of the respective propagators, dropping all (irrelevant)
terms proportional to $p_\m p_\n$:  That of each TME is the sum of
an ordinary (even) Proca term, $\eta_{\mu\nu}[p^2 + m^2]^{-1}$, and
a mass- and parity-odd one $\sim (m \, \e_{\mu\nu\a}p^\a ) p^{-2}
(p^2 + m^2)^{-1}$, which cancels out in the $\pm m $ doublet.  Hence
the
interaction is of the usual finite-range form $\sim j^\m  (p^2 + m^2)^{-
1}j_\m$ in both theories.

It is instructive to establish how equivalence is displayed at the level
of the field equations, and especially how the manifest TME gauge
invariance can be ``hidden" in the process.  The field equations are
$$ %3a
\pa_\m F^{\mu\nu}(A_1) + m~^*F^\n (A_1) =
\textstyle{\frac{1}{\sqrt{2}}} \: j^\n
\eqno{(3{\rm a})}
$$
$$ %3b
\pa_\m F^{\mu\nu}(A_2) - m~^*F^\n (A_2) =
\textstyle{\frac{1}{\sqrt{2}}} \: j^\n \; .
\eqno{(3{\rm b})}
$$
Here $^*F^{\nu} \equiv \frac{1}{2} \, \e^{\n\a\b}F_{\a\b}$ is the
dual
field strength, with $F_{\mu\nu} = - \e_{\mu\nu\a}~^*F^\a$.  In
terms of the sums and differences,
$A^\pm_\m = \frac{1}{\sqrt{2}} \, (A^1_\m \pm A^2_\m )$, we have
$$
\pa_\m \, F^{\mu\nu} (A_+) + m \, ^*F^\n (A_-) = j^\n
\eqno{(4{\rm a})}
$$
$$
\pa_\m \, F^{\mu\nu} (A_-) + m \, ^*F^\n (A_+)
\equiv \e^{\n\m\a} \pa_\m [^*F_\a (A_-) + mA^+_\a ] = 0 \; .
\eqno{(4{\rm b})}
$$
The general solution of (4b) is
\renewcommand{\theequation}{\arabic{equation}}
\setcounter{equation}{4}
\begin{equation}%5
 ^*F_\a (A^-) + m \: A^+_\a = \pa_\a \L + F_\a (V)
\end{equation}
where $V_\m$ is a solution of the homogeneous Maxwell equation,
$\e^{\mu\nu\a} \e^{\a\l\s} \pa^2_{\n\l} V_\s = 0$.  Inserting (5)
into (4a) yields the Proca equation in (not quite) Stueckelberg form,
\begin{equation}%6
\pa_\m F^{\mu\nu}(A_+) - m^2
[A^\n_+ - m^{-1} \pa^\n \L -m^{-1}~^*F^\n (V)] = j^\n \; .
\end{equation}
The gauge-freedom carried by $\L$ can be field-redefined away by
$A^\n_+ \rightarrow A^\n_+ - m^{-1} \pa^\n \L$ as usual.  But so
can the additional $V$ term, by $A^\n_+ \rightarrow A^\n_+ - m^{-1}
\,^*F^\n (V)$ since
$\pa_\m \,F^{\mu\nu}(^*F (V))$ is just $\Box~^*F^\n (V)$, which
vanishes for a field
strength
$^*F^\n(V)$ obeying Maxwell's equations.  So the $A_+$ field's
invariance is absorbed by an extended Stueckelberg transformation,
leaving the Proca form.  Once the $A_+$-field is determined (up to this
gauge) by (6), we can use (4b) to determine $A_-$ (up to a gauge), with
$m^*F^\n (A_+)$ as the effective current determining
$F^{\mu\nu}(A_-)$ through an ordinary massless Maxwell equation.

The equivalence of the two systems' degrees of freedom  goes as
follows.  Each TME embodies a single massive particle with helicity
$m/|m|$.  Proca theory obviously has
(D--1) = 2 massive degrees of freedom, and they too have
helicities $\pm$1, though this fact requires the same analysis of the
full
Lorentz
algebra, (rather than merely that of the rotation generator $M^{ij}$),
as was required for TME \cite{002}; as we shall see, the naive $M^{ij}$
is purely
orbital for both
systems.  For completeness, we identify the excitations in terms of the
canonical formulation of both
models, using first-order form of (1),(2) where $(A_\a , \; F^{\mu\nu})$
are initially independent variables.  The free Lagrangian
\begin{equation}%7
L = -\,\textstyle{\frac{1}{2}}\, F^{\mu\nu}(\pa_\m A_\n -
\pa_\n A_\m ) + \textstyle{\frac{1}{4}} \, F^{\mu\nu}F^{\a\b}
\eta_{\mu\a}\eta_{\nu\b} + \textstyle{\frac{\a}{2}} \, m \:
\e^{\mu\a\nu}A_\m\pa_\a A_\n - \textstyle{\frac{\b}{2}} \,
m^2 \, A^2_\m
\end{equation}
represents TME or Proca theory, as $\a = 1, \; \b = 0$ or $ \a = 0, \;
\b = 1$ respectively.\footnote{Keeping both terms results in a
parity-violating theory with two different helicities, as does taking the
CS term to be of the form
$\e^{\a\mu\nu}A_\a F^{\l\s}\eta_{\mu\l}\eta_{\nu\s}$, which is no
longer
gauge-invariant, nor metric-independent.}  The stress tensor associated
to (7)
is the usual (massive) vector one,
\begin{equation}%8
T_{\mu\nu} = F_\m\!~^{\a} F_{\nu\a} - \textstyle{\frac{1}{4}}\,
\eta_{\mu\nu} F^2_{\a\b} + \b m^2 (A_\m A_\n -
\textstyle{\frac{1}{2}} \, \eta_{\mu\nu} A^2_\a ) \; ;
\end{equation}
the CS term (being metric-independent) does not contribute.  Let us
briefly recapitulate the canonical reduction of (7) in terms of the
space-time decomposition of its variables.  Our conventions and
notations are
$\e^{012} = \e^{12} = +1 = - \e_{012}$,
\begin{equation}%9
F^{0i} = F_{i0} = E^i \; , \;\;\;\; B \equiv \textstyle{\frac{1}{2}} \,
\e^{ij} F_{ij} \; ,
\end{equation}
while the transverse-longitudinal decomposition of a spatial vector is
expressed by
\begin{equation}%10
V^i = V^i_T + V^i_L = \e^{ij}\xi^{-1}\pa_jv_T +
\xi^{-1} \pa_i v_L \; , \;\;\; \xi = (-\nabla^2)^{1/2} \; .
\end{equation}
In both theories, one may eliminate the trivial constraint
\begin{equation}%11
F_{ij} = \pa_i A_j - \pa_j \, A_i = \e^{ij} B =
\e^{ij} \xi a_T \; .
\end{equation}
However, in TME, only the gauge-independent pair $(e_T , \:a_T)$
remains, whereas the longitudinal components also survive in Proca
theory.  The difference lies in the fact that $A_0$ is respectively a
Lagrange multiplier or an auxiliary field in the two systems; that is, its
variation yields the constraint
\begin{equation}%12
- \nabla \! \cdot \! E + \textstyle{\frac{\a}{2}} \, m B + \b m^2 A_0
= j^0
\; .
\end{equation}
Dropping the source for the moment, we then find for the free actions,
which began as
\begin{eqnarray}%13
I = & - & \int d^3x [\mbox{\boldmath $E$}\cdot
(\dot{\mbox{\boldmath $A$}} - \mbox{\boldmath $\nabla$} A_0) +
\textstyle{\frac{1}{2}} \,
\mbox{\boldmath $E$}^2 + \textstyle{\frac{1}{2}} B^2 (A) +
\a m \,
\{ \e^{ij} A_i \dot{A}_j - \textstyle{\frac{1}{2}\, A_0 B(A)} \}
\nonumber \\[.1in]
& + & \textstyle{\frac{\b}{2}} \, m^2 (\mbox{\boldmath $A$}^2 -
A^2_0 )] \; ,
\end{eqnarray}
the final forms
\begin{eqnarray}%14
I_{\rm TME} & = & \int d^3x [(-e_T) \dot{a}_T - \textstyle{\frac{1}{2}}
\,
e^2_T - \textstyle{\frac{1}{2}} \, a_T (m^2-\nabla^2)a_T] \; ,
\\ [.1in]
I_P & = & I_{\rm TME} + \int d^3x
[\bar{a}_L \dot{\bar{e}}_L
- \textstyle{\frac{1}{2}} \, \bar{a}^2_L
- \textstyle{\frac{1}{2}} \bar{e}_L
(m^2 - \nabla^2)\bar{e}_L ] \; .
\end{eqnarray}
The doublet action is just the sum of two independent terms (14).  To
reach (14), we eliminated the longitudinal electric field using the Gauss
constraint (12).  To obtain (15), we used (12) for $\a = 0$, $\b = 1$, to
eliminate $A_0$, and rescaled $\bar{a}_L \equiv m^{-1}a_L$,
$\bar{e}_L
\equiv m e_L$.  The free field content then, is that both theories have
two massive degrees of freedom.  To determine their helicities requires,
in both cases, study of the full Lorentz algebra, for the
rotation generators are ostensibly purely orbital  because of the
peculiarities of
2--space.  That is,
from its definition,
\begin{equation}%16
M = \int d^2r \, \e_{ij} \, x^i \, T^j_0 \; ,
\end{equation}
it follows from (8) and the canonical reductions that
$$%17a
M = \int d^2r \sum^2_{i = 1} (-e^i_T) \pa_\th a^i_T
\eqno{(17{\rm a})}
$$
for the TME doublet and
$$
M = \int d^2r [(-e_T)\pa_\th a_T
+ \bar{a}_L \pa_\th \bar{e}_L ]
\eqno{(17{\rm b})}
$$
for the massive vector, as is to be expected since the canonical variables
are spatial scalars.

We now reinstate the $j^\m A_\m$ term to recover the effective
current-current coupling in the present context.  The canonical TME
doublet action (14) contains a term $\sim \textstyle{\frac{1}{\sqrt{2}}}
j_T
(a^1_T + a^2_T)$ together with the longitudinal interactions, which for
our conserved currents would look like
$$
\rho \, \frac{\Box}{(m^2-\Box )} \, \nabla^{-2}\rho \sim
\rho (-\nabla^{-2})\rho +
\rho \, \frac{(m^2/\nabla^2)}{m^2 - \Box} \, \rho \; .
$$
This is the usual sum of a Coulomb term and a (retarded) Yukawa
interaction, whereas the $j_T a_T$ coupling above results in the purely
retarded form
$j_T (m^2 - \Box )^{-1} j_T$.   The Proca action would
have, in addition to the $j^T a_T$ and $j^L a_L$ terms, the
remnant of $\rho A_0$, which together of course yield the same $\rho
- \rho$ interactions as above for TME doublets.  Hence the total
current-current coupling is of course the same $j^\m (m^2 - \Box )^{-
1} j_\m$ in both cases, and limits smoothly to the Maxwell one as $m
\rightarrow 0$.

\section{Tensor Theories}

The gravitational case is similar to the vector in most respects.  A
doublet of opposite mass TMG has the same excitation content as a
single massive, Pauli--Fierz, tensor theory and leads to essentially
equivalent induced source-source interactions \cite{002}.

The doublet's action is \cite{002}
\renewcommand{\theequation}{\arabic{equation}}
\setcounter{equation}{17}
\begin{eqnarray}%18
I_{TD} & = & I_{\rm TMG} ( h_{\m\nu}^1 ,\m ) + I_{\rm TMG}
(h_{\m\nu}^2 ,-\m ) +
\textstyle{\frac{1}{\sqrt{2}}} \, \k T^{\mu\nu} (h^1_{\m\nu} +
h^2_{\mu\nu})
\nonumber \\ [.1in]
I_{\rm TMG} (h_{\m\nu},\m ) & = & - \k^{-2} \int d^3x  R^Q (h) +
1/2\m \int d^3x \e^{\mu\a\b}G^\n_\a \pa_\m (\k h_{\b\n}) \; ,
\end{eqnarray}
where $Q$ indicates that only terms quadratic in $\k h_{\mu\nu}
\equiv (g_{\mu\nu} - \eta_{\m\n} )$ are to be kept in the expansion
of the
Einstein actions and $G^\n_\a$ is linearized.  Again, parity
conservation is restored by including the interchange
$h^1_{\mu\nu} \leftrightarrow h^2_{\mu\nu}$.  The sign of the
Einstein term is ``ghost-like," {\it i.e.}, opposite of the conventional one,
in order that the TMG excitations be non-ghost.  The Einstein constant
$\k^2$ has dimensions of length, while $\m$ is dimensionless since
the CS term is of third derivative order; the mass $m$ is
$\m \k^{-2}$, but the ``massless" Einstein limit is clearly $\m
\rightarrow \infty$.  The Pauli--Fierz action is as in D=4,
\begin{equation}%19
I_{PF} = + \int d^3x R^Q (k) - \frac{m^2}{2} \int d^3x
(k^2_{\mu\nu} - k^2 ) + \k \int d^3x k_{\mu\nu}
T^{\mu\nu} \; , \;\;\; k \equiv k^\a_\a \; .
\end{equation}
The couplings in both cases are the usual minimal ones,
and $T^{\mu\nu}$ is necessarily conserved for consistency with the
(linearized) gauge invariance of TMG; henceforth we
assume\footnote{Of course, just as in normal general relativity, a
dynamical $T_{\mu\nu}$ will no longer be conserved as a result of its
coupling and the full nonlinear theory will be required.} $\pa_\mu
T^{\mu\nu} = 0$.  The propagator of a single TMG field (neglecting
terms proportional to $p^\m p^\n$, which vanish for conserved
sources) differs from that of the massive one in two respects:  it
contains, in addition, a term with numerator
$\sim \m (\e^{\m\a\b}p^\g +$ symm) and also a term independent
of
$\m$, corresponding to a free (ghost) Einstein field propagator.
In the doublet then, the odd terms cancel, leaving two interaction
terms,
the usual Pauli--Fierz finite range one,
$$%20a
\k^2 \int d^3x [T^{\mu\nu}(-\Box + m^2)^{-1} T_{\mu\nu} -
\textstyle{\frac{1}{2}} \, T^\m_\m (-\Box +m^2 )^{-1} T^\n_\n ]
\eqno{(20{\rm a})}
$$
and the extra ghost Einstein part,
$$%20b
-\k^2 \int d^3x [T^{\mu\nu}(-\Box )^{-1} T_{\mu\nu} -
 T^\m_\m (-\Box )^{-1} T^\n_\n ] \; .
\eqno{(20{\rm b})}
$$
[The coefficients of the $T^\m_\m T^\n_\n$ terms are (D--1)$^{-1}$
and (D--2)$^{-1}$ respectively for general D.]  The Einstein ``interaction"
is, however
an artifact, consisting entirely  of contact terms when conservation of
$T^{\mu\nu}$ is taken into account \cite{011}.  The  finite-range
interaction (20a) is then effectively the entire residue in both systems,
and equivalence of interactions is established modulo the (trivial)
Einstein part (20b) in the doublet.

Now let us analyze the field equations of the TMG doublet, as we did
for TME.  They are
$$%21a
-G_{\mu\nu} (h_1 ) + m^{-1} C_{\mu\nu} (h_1 ) =
\textstyle{\frac{1}{\sqrt{2}}}\,\k
\, T_{\mu\nu}
\eqno{(21{\rm a})}
$$
$$%21b
-G_{\mu\nu} (h_2 ) - m^{-1} C_{\mu\nu} (h_2 ) =
\textstyle{\frac{1}{\sqrt{2}}}\,\k\,
T_{\mu\nu} \; ,
\eqno{(21{\rm b})}
$$
where $G_{\mu\nu}$ is the linearized Einstein tensor and
$C_{\mu\nu}$ is the third-derivative order linearized Cotton--Weyl
tensor
$C^{\mu\nu} \equiv \e^{\m\a\b} \pa_a (R^\n_\b -
\textstyle{\frac{1}{4}} \, \d^\n_\b \, R)$; it is symmetric, traceless and
(identically) conserved.  We again take the sum and difference of (21a,b)
in terms of
$h^\pm_{\mu\nu} = \frac{1}{\sqrt{2}} (h^1_{\mu\nu} \pm
h^2_{\mu\nu})$,
$$%22a
-G_{\mu\nu} (h^+) + m^{-1} C_{\mu\nu} (h^-) = \k \, T_{\mu\nu}
\eqno{(22{\rm a})}
$$
$$%22b
-G_{\mu\nu} (h^-) + m^{-1} C_{\mu\nu} (h^+) = 0 \; .
\eqno{(22{\rm b})}
$$
One may solve (22b) for either $h^-_{\mu\nu}$ or $h^+_{\mu\nu}$
and insert into (22a) to get an equation in terms of the other; this will
yield third or fourth derivative inhomogeneous equations.  I will sketch
the results, but to save space will omit all the ``Stueckelberg" gauge
parts as well as required symmetrizations that arise in the process; I
will also drop scalar curvatures since it is the Ricci tensor that counts
(actually the trace of (22b) shows that $R(h^-)$ vanishes).  All
``equalities" below that are subject to these caveats will carry $\approx$
signs.  Using the fact that
$G^{\mu\nu}(h) =  - \frac{1}{2} \, \e^{\m\a\b} \e^{\n\l\s}
\pa^2_{\a\l} h_{\b\s}$, we learn from (22b) that, in the obvious
notation
$R^+_{\n\b} \equiv R_{\n\b} (h^+)$,
\renewcommand{\theequation}{\arabic{equation}}
\setcounter{equation}{22}
\begin{equation}%23
\e^{\m\a\b} \pa_\a (m^{-1} R^+_{\n\b} + \textstyle{\frac{1}{2}} \,
\e^{\n\l\s} \pa_\l \, h^-_{\b\s} ) \approx 0 \; .
\end{equation}
Apart from homogeneous (gauge) terms, then, we may eliminate
$R^+_{\nu\b}$ in terms of first derivatives of $h^-_{\b\s}$ in (22a) to
find the
third order equation
$$%24a
\e^{\mu\l\s} \pa_\l ( 2 R^-_{\s\n} + m^2 h^-_{\s\n} ) \approx 2 m\k
\, T_{\mu\nu} \; .
\eqno{(24{\rm a})}
$$
The left side is the gauge-invariant curl of a Klein--Gordon form; in
harmonic gauge where
$2 R^-_{\s\n} = - \Box h^-_{\s\n}$, we have
$$%24b
\e^{\mu\l\s} \pa_\l (-\Box + m^2 ) h^-_{\s\n} \approx 2 m\k \,
T_{\mu\nu} \; .
\eqno{(24{\rm b})}
$$
Alternatively, we may take the curl of (22b) to learn that
$mC^-_{\mu\nu} = \Box R^+_{\mu\nu}$, and hence we may write
(22a) as
$$%25a
(m^2 - \Box ) R^+_{\mu\nu} \approx  -\k m^2 \, T_{\mu\nu} \; .
\eqno{(25{\rm a})}
$$
This is a (fourth order) Klein--Gordon equation which states that, in
harmonic gauge,
$$%25b
\Box (m^2 - \Box ) h^+_{\mu\nu} \approx  2 \k m^2 \, T_{\mu\nu}
\; .
\eqno{(25{\rm b})}
$$
[Of course (24b) and (25b) just reflect the propagator structure of TMG
and its doublets that was given earlier.]

Let us now compare the above TMG equations with the Pauli--Fierz
ones,
\renewcommand{\theequation}{\arabic{equation}}
\setcounter{equation}{25}
\begin{equation}%26
G_{\mu\nu}(k) + \textstyle{\frac{m^2}{2}} (k_{\mu\nu} -
\eta_{\mu\nu} \, k) = \k\, T_{\mu\nu} \; .
\end{equation}
The analysis goes as in D=4, namely one
first notes that (for conserved $T_{\mu\nu}$) the divergence of (26)
implies
$\pa^\m (\eta_{\mu\nu} \, k - k_{\mu\nu} ) = 0$; the double
divergence is
just the scalar curvature $R(k)$, which then also vanishes.  Using these
constraints, we may write $G_{\mu\nu}(k)$ as follows:
\renewcommand{\theequation}{\arabic{equation}}
\setcounter{equation}{26}
\begin{eqnarray}%27
G_{\mu\nu}(k) \; = \; R_{\mu\nu}(k) & = & - \textstyle{\frac{1}{2}}
(\Box k_{\mu\nu} - \pa^2_{\a\m} k^\a_\n - \pa^2_{\a\n}
k^\a_\m + \pa^2_{\mu\nu} k) \nonumber \\[.1in]
& = & - \textstyle{\frac{1}{2}} (\Box k_{\mu\nu} - \pa^2_{\mu\nu} k)
\; .
\end{eqnarray}
Now perform the ``Stueckelberg" field redefinition,
$k_{\mu\nu} \rightarrow k_{\mu\nu} - m^{-2} \pa^2_{\mu\nu} k$ in
(26), under which $G_{\mu\nu}$ is of course invariant.  This change in
form of the mass term just cancels the $\pa^2_{\mu\nu}k$ in (27) and
yields the standard Klein--Gordon equation
\begin{equation}%28
(m^2 - \Box ) (k_{\mu\nu} - \eta_{\mu\nu} k) =
2 \k T_{\mu\nu} \; .
\end{equation}
Comparing with the corresponding (schematic!) equations for the TMG
doublet, (24b) or (25b), we see there the same basic Klein--Gordon
propagation, modified by the higher order character of TMG.  Had we
kept the $R^+$ terms, the gauge parts, performed Stueckelberg shifts,
etc., we would have again deduced from these field equations the
equivalence of source-source interactions modulo the extra Einstein
``coupling" in TMG.

We will not reproduce the canonical analysis of TMG, which may be
found in \cite{002}, nor that of the massive theory, well-known from
D=4.  The pattern is clear from the vector story:  in both cases there
are two massive excitations, whose helicity $\pm$2 character cannot be
determined from the rotation generators alone.

\section{Differences}

The correspondence between the TM gauge doublets and their normally
massive counterparts is not complete.  We discuss three types of
differences here.  The first deals with existence of long-range
Coulomb-like (but locally pure gauge) potentials in the gauge doublets
--- despite the finite
range character of the excitations and interactions --- this is the most
manifest aspect of their gauge invariance, with no massive counterpart.
Secondly, we analyse the zero mass limits of the free excitation spectra,
and thirdly, that of the source-source interactions.  The latter aspect
displays particularly striking differences in the tensor case:  TMG leads
to a smooth (and hence trivial) Einstein limit, whereas the Pauli--Fierz
discontinuity of D=4 \cite{009} not only persists in D=3 but now gives
a non-trivial, and therefore especially ``different," interaction from that
of Einstein gravity.

The Proca and Pauli--Fierz equations consist entirely of a set of
Klein--Gordon equations of the form $(m^2 - \Box )\phi_i = \rho_i$,
with no differential constraints and hence no long-range components.
The TM theories, however, do contain Gauss constraints, but with
additional terms.  In the vector case, (12) shows (for $\b = 0$) that it
is now $mB$, rather than the short-range $\nabla \! \cdot \! E$ which
carries the asymptotic information \cite{002}, because $mB$ is a curl
whose spatial integral is proportional to the total charge
$Q = \int d^2 r\, j^0$.  We therefore expect that in our doublet, it is
the difference, $B (A_-)$, which is relevant.  This is indeed the case:
while the $A^+_\m$-field is short-range, obeying as it does the Proca
equation (6), it follows from (5) that $A^-_\m$ is determined then by
$A^+$.  Specifically, $^*F^0(A_-) = B(A_-)$ is proportional to the sum
of $\nabla \! \cdot \! E(A_+)$ and the charge density $j^0$.
Consequently,
since $E(A_+)$ is short-range, it is $A_-$ at spatial infinity that is
proportional to $Q = \int d^2 r \, j^0$.  This can also be read off from
the time component of (5): the right hand side's spatial integral
vanishes, while the integral of $A^+_0$ is proportional to $Q$, by the
Proca equation.

A very similar picture emerges in the tensor case:  the lower derivative
part, $G^0_{0}$, of the TMG field equations (21) carries the asymptotic
information about the energy of the source \cite{002}, just as it does in
Einstein gravity.  But this time it is the term even in $m$, and hence
$G^0_{0}(h^+)$ that is relevant.  Now $G^0_{0}(h^+)$ is (like $B(A)$)
a total spatial divergence, and the source energy is the spatial integral
of $T^0_{0}$, so we need only integrate the (00) components of (22a)
over space, noting that the $C^-$ term falls off too rapidly to contribute
at infinity.

Next we compare the massive and massless excitation spectra, using
the well-known fact that for each gauge symmetry broken by a lower
derivative term, one new degree of freedom arises (only one, because
the former Lagrange multiplier, unlike the gauge variable, is merely
promoted to be an auxiliary field).  Thus massive vector theory
acquires one degree of freedom beyond the (D--2) of Maxwell theory
because $m^2A^2_\m$ breaks its gauge invariance.  The massive
tensor
model has (D--1) additional excitations beyond the $\frac{1}{2}$ D(D--3)
of Einstein theory because $m^2 (h^2_{\mu\nu} - h^2)$ breaks all but
the longitudinal one of the D gauges of linearized gravity; so there is
always a discontinuity (except at D=2, where the Einstein kinetic term
is absent altogether).  On the other hand, in (singlet) TME, the CS
term shares
the Maxwell invariance and there is no discontinuity in the number,
but only in the helicity, of the single excitation (massless particles
always have  vanishing helicity \cite{012}).  [This method of counting
also applies in
vector theory when a ``mass" term is added to a pure CS term
\cite{005}, breaking the latter's gauge invariance to give rise to a single
massive degree of freedom; it is in fact equivalent to TME \cite{006},
as is also the case if the mass term arises through a Higgs mechanism
\cite{007}.]  In TMG, while the Einstein term also preserves
gauge invariance of the CS term, it breaks the latter's conformal
invariance, thereby generating the massive graviton discontinuously
from its two separately nondynamical parts.

More dramatic are the differences between source-source interactions
in our models and in their $m=0$ or $m=\infty$ counterparts. Whereas
the degree of freedom count is a more formal difference -- for example,
it has long been known from D=4 that the extra degree of freedom in
Proca theory decouples from the current as $m \rightarrow 0$, so that
the limit is (all but gravitationally) indistinguishable from
electrodynamics, there is a discontinuity in the induced source couplings
already in D=4 between Einstein and Pauli--Fierz models \cite{005},
precisely because not all the extra modes decouple from the stress
tensor in the massless limit.  Here TMG provides a clear difference
from the massive model:
As we have seen by considering the propagator, the infinite mass
limit\footnote{For completeness, we note that the $m\rightarrow 0$
limit of TMG is to be contrasted with pure CS, rather than Einstein,
gravity.  Here our doublet is now ``discontinuous,"  since it leads to a
net pure trace, $\sim T^\m_\m \Box^{-1} T^\n_\n$ coupling, whereas
pure CS gravity of course cannot even couple to sources whose traces
fail to vanish; continuity is of course restored if the source is traceless.
The Pauli--Fierz theory (like Proca theory) of course
leaves no interactions at all in the infinite mass limit.} of the TMG
doublet (where
its free action reduces to the Einstein part) leads to just the same
$T_{\mu\nu} - T_{\mu\nu}$ coupling (or rather lack of coupling!) as in
Einstein theory.  On the other hand, the massless (``Einstein") limit of
Pauli--Fierz coupling differs from the Einstein form by the famous
(D--1)$^{-1}$ versus (D--2)$^{-1}$ coefficient of the
$-T^\m_\m \Box^{-1} T^\n_\n$ term of (20).  This is especially
dramatic in D=3 where it means in particular the difference between,
respectively,
existence or absence of a Newtonian limit!

\section{Summary}

We have reviewed the resemblances and differences between the two
ways of giving mass to gauge fields in D=3, through gauge-invariant
Chern--Simons terms or through explicit mass terms.  The equivalences
between these ways, despite their different gauge properties, were
followed also at the level of field equations where the ``fading" of gauge
invariance into ordinary Klein--Gordon-like expressions was understood
as a manifestation of Stueckelberg mechanisms.  Most dramatic among
the differences was the fact that TMG provides the only example of a
smooth limit for the interactions from massive to Einstein form, in
contrast to the discontinuity in the explicitly massive tensor theory.

\end{document}